\documentclass[lettersize,journal]{IEEEtran}
\usepackage{amsmath,amsfonts,amsthm}

\usepackage{algorithmic}
\usepackage{algorithm}
\usepackage{array}
\usepackage{subfigure}
\usepackage{textcomp}
\usepackage{stfloats}
\usepackage{url}
\usepackage{verbatim}
\usepackage{graphicx}
\graphicspath{{image/}}
\usepackage{cite}
\usepackage{booktabs}
\begin{document}

\title{MambaLithium: Selective state space model for remaining-useful-life, state-of-health, and state-of-charge estimation of lithium-ion batteries}

\author{Zhuangwei Shi
\thanks{Correspondence author: Zhuangwei Shi (email:zwshi@mail.nankai.edu.cn)}
\thanks{Zhuangwei Shi is with the Ji Hua Laboratory, Foshan, Guangdong, China.}}

\markboth{Journal of \LaTeX\ Class Files,~Vol.~14, No.~8, August~2021}%
{Shell \MakeLowercase{\textit{et al.}}: A Sample Article Using IEEEtran.cls for IEEE Journals}


\maketitle

\begin{abstract}
    Recently, lithium-ion batteries occupy a pivotal position in the realm of electric vehicles and the burgeoning new energy industry. Their performance is heavily dependent on three core states: remaining-useful-life (RUL), state-of-health (SOH), and state-of-charge (SOC). Given the remarkable success of Mamba (Structured state space sequence models with selection mechanism and scan module, S6) in sequence modeling tasks, this paper introduces MambaLithium, a selective state space model tailored for precise estimation of these critical battery states. Leveraging Mamba algorithms, MambaLithium adeptly captures the intricate aging and charging dynamics of lithium-ion batteries. By focusing on pivotal states within the battery's operational envelope, MambaLithium not only enhances estimation accuracy but also maintains computational robustness. Experiments conducted using real-world battery data have validated the model's superiority in predicting battery health and performance metrics, surpassing current methods. The proposed MambaLithium framework is potential for applications in advancing battery management systems and fostering sustainable energy storage solutions. Source code is available at \url{https://github.com/zshicode/MambaLithium}.
\end{abstract}

\begin{IEEEkeywords}
    Mamba, selective state space model, remaining-useful-life (RUL), state-of-health (SOH) and state-of-charge (SOC)
\end{IEEEkeywords}

\section{Introduction}
\label{sec:introduction}

\IEEEPARstart{I}{n} recent years, lithium-ion batteries have emerged as the preferred energy storage solution in various applications ranging from electric vehicles to portable electronics. However, the accurate estimation of their remaining-useful-life (RUL), state-of-health (SOH), and state-of-charge (SOC) remains a challenging task due to the complex electrochemical processes involved and the various factors that affect battery performance. Accurate estimation of these parameters is crucial for effective battery management, optimization of energy usage, and ensuring the safety and reliability of battery-powered systems.

Existing methods for RUL, SOH, and SOC estimation often rely on complex electrochemical models or data-driven approaches. While electrochemical models provide a detailed understanding of battery behavior, they can be computationally intensive and difficult to calibrate accurately. As RUL, SOH, and SOC estimation can be modeled as time series forecasting problems, data-driven approaches such as machine learning and deep learning, have shown promise in handling the nonlinearities and uncertainties inherent in battery systems. However, these methods often require large amounts of labeled data for training, which can be challenging to obtain in practical scenarios.

The traditional time series model ARIMA (Autoregressive Integrated Moving Average Model) \cite{boxjenkins1970,ZHANG2003159} can not describe the nonlinear time series, and needs to satisfy many preconditions before modeling, and can not achieve remarkable results in the stock forecasting. In recent years, with the rapid development of artificial intelligence theory and technology, more and more researchers apply artificial intelligence method to the financial market. On the other hand, the sequence modeling problem, focusing on natural language sequences, protein sequences, stock price sequences, and so on, is important in the field of artificial intelligence research \cite{shi2021vgaelda,jin2021lpigac}. The most representative artificial intelligence method is neural networks, which are with strong nonlinear generalization ability.

Recurrent Neural Network (RNN) was adopted for analyzing sequential data via neural network architecture, and Long Short-Term Memory (LSTM) model is the most commonly used RNN. LSTM introduced gate mechanism in RNN, which can be seen as simulation for human memory, that human can remember useful information and forget useless information \cite{jin2021chinese,jin2022tlcrys}. Attention Mechanism \cite{transformer2017} can be seen as simulation for human attention, that human can pay attention to useful information and ignore useless information. Attention-based Convolutional Neural Networks (ACNN) are widely used for sequence modeling \cite{JIN2021265,lin2022dattprot} and complex dependency captureing \cite{jin2022nimgsa,shi2022graph}. Combining Attention-based Convolutional Neural Networks and Long Short-Term Memory, is a self-attention based sequence-to-sequence (seq2seq) \cite{seq2seq2014} model to encode and decode sequential data. This model can solve long-term dependency problem in LSTM, hence, it can better model long sequences. LSTM can capture particular long-distance correspondence that fits the sturcture of LSTM itself, while ACNN can capture both local and global correspondence. Therefore, this architecture is more flexible and robust.

Transformer \cite{transformer2017} is the most successful sequential learning self-attention based model. Experiments on natural language processing demonstrates that Transformer can better model long sequences. Bidirectional Encoder Representation Transformer (BERT) with pretraining \cite{Devlin2018BERT} can perform better than the basic Transformer. Pretraining  is a method to significantly improve the performance of Transformer (BERT). Attention-based neural networks are widely used in a hybrid framework for time series forecasting as a key module. AttCLX \cite{shi2022attclx} proposed an attention-based CNN-LSTM and XGBoost hybrid model for stock prediction. Recently, neural network-enhanced state space Kalman Filter models have been applied for time-series forecasting. TL-KF \cite{shi2021tlkf} proposed Kalman Filter along with LSTM and Transformer.

The time series forecasting models mentioned above have been adopted for RUL, SOH, and SOC estimation. Yang et al. \cite{yang2019soc} introduced an LSTM-based approach for accurate state-of-charge estimation of lithium-ion batteries. Kong et al. \cite{kong2021voltage} proposed a method that utilizes voltage-temperature health features to enhance prognostics and health management of lithium-ion batteries. Wen et al. \cite{wen2024physics} presented physics-informed neural networks for prognostics and health management of lithium-ion batteries. Chen et al. \cite{chen2024lstm} developed an LSTM-SA model that estimates the state-of-charge of lithium-ion batteries under varying temperatures and aging levels. However, the improvement of the effectiveness in handling the complex nonlinearities and time-varying characteristics of battery systems is still required.

The Mamba model \cite{gu2023mamba} represents a significant advancement in the field of sequence modeling. This model outperforms the traditional approach by incorporating a structured state space sequence model (S4) with a selection mechanism and scan module, known as S6. The Mamba model excels at capturing nonlinear patterns in sequential data, which has historically been a challenge for traditional time series models. The core strength of Mamba lies in its ability to efficiently model sequences using a linear-time complexity, making it suitable for processing large-scale datasets. The innovative selection mechanism allows it to dynamically adapt to different patterns and structures within the data, enabling more accurate predictions. Additionally, the scan module enhances the capability by scanning through the state spaces to identify relevant information for making predictions. 

Due to its versatility and adaptability, Mamba have been a popular choice for various sequence modeling tasks \cite{gu2023mamba,shi2024mamba}. However, the application of Mamba in time series of battery systems remains to be explored. Therefore, this paper proposes a Mamba-based model for RUL, SOH and SOC estimation of lithium-ion batteries, named MambaLithium. By harnessing the powerful algorithms of Mamba, MambaLithium skillfully captures the intricate aging and charging behaviors of lithium-ion batteries. By zeroing in on the crucial states within the operational range of the battery, MambaLithium not only boosts estimation precision but also ensures computational stability. Rigorous experiments using real-world battery data have confirmed the model's preeminence in predicting battery health and performance indicators, outperforming existing techniques. The proposed MambaLithium framework holds promise for enhancing battery management systems and promoting sustainable energy storage solutions. The source code of this paper is available at \url{https://github.com/zshicode/MambaLithium}.

\section{Materials and Methods}

\subsection{RUL, SOH and SOC of lithium-ion batteries} \label{sec:lithium}

The charging-discharging cycles of lithium-ion batteries is limited. At the $k$-th cycle, the remaining-useful-life $RUL_k$ defines the number of remaining cycles that a battery can be used. The capacity of battery decreases as the used cycles increases. Let $Q_k$ denotes the capacity at the $k$-th cycle, and $Q_N$ denotes the nominal capacity, the state-of-health is defined as
\begin{equation}
    SOH_k=\frac{Q_k}{Q_N}
\end{equation}

In the charging procedure of a certain cycle, let $Q_i$ (A$\cdots$ h) and $V_i$ (V) denotes the capacity and voltage at time $i$ respectively. The $Q-V$ curve is called incremented capacity (IC) curve. Following \cite{kong2021voltage,wen2024physics},

\begin{equation}
    Q_{i+1}-Q_i=-\omega Q_i^2+b+\epsilon
\end{equation}

Here, $\omega,b$ are two parameters fitted by the regression of IC curve, and $\epsilon$ denotes the Gaussian noise. $dQ/dV$ can be calculated from the slope of IC curve. Besides, the average temperature $T$ (${}^\circ$C) during the charging procedure, the internal resistance $r$ ($\Omega$) and the charge time $\tau$ (h) are also important factors. For RUL and SOH prediction, the feature of $k$-th cycle includes $\omega_k,b_k,T_k,r_k,\tau_k,(dQ/dV)^{max}_k,(dQ/dV)^{min}_k,\mathrm{Var}[dQ/dV]_k$. 

In the discharging procedure of a certain cycle, let $Q_i$ (A$\cdots$ h) and $I_i$ (A) denotes the capacity and current at time $i$ respectively, the state-of-charge is defined as
\begin{equation}
    SOC_i=SOC_0-\frac{1}{Q_N}\int_0^i I_tdt
\end{equation}

For RUL and SOH prediction, the feature of time $i$ cycle includes $I_i,V_i,T_i$.

\subsection{Structured state space sequence model (S4)}

State space model is inspired by solving ODE \cite{kalman1960,shi2022diff}. Structured State Space sequence model (S4) is a recently proposed sequence modeling architecture that leverages the power of state spaces and structured matrices for efficient and effective processing of sequential data. It combines principles from control theory, signal processing, and deep learning to address the challenges associated with traditional sequence modeling approaches.

In S4 model, the key idea is to represent the underlying dynamics of a sequence using a state space representation. This representation captures the evolution of the systematic state over time, allowing for efficient computation and storage. The state space is parameterized using structured matrices, which impose certain constraints on the parameters to enable efficient training and inference.

Let $X\in\mathbb{R}^{B\times L\times D}$, where $B,L,D$ denote batch size, time steps and dimension, respectively. For each batch and each dimension, let $x_t,h_t,y_t$ denote the input, hidden state and output at time $t=1,2,...,L$, respectively, S4 model can be written as
\begin{equation}
    h_t=Ah_{t-1}+Bx_t
\end{equation}
\begin{equation}
    y_t=Ch_t
\end{equation}

By discretization, let $\Delta$ denotes the sample time, then
\begin{equation}
    \bar A=\exp(\Delta A)
\end{equation}
\begin{equation}
    \bar B=(\Delta A)^{-1}(\exp(\Delta A)-I)\cdot\Delta B
\end{equation}
\begin{equation}
    h_t=\bar Ah_{t-1}+\bar Bx_t
\end{equation}

Here, $A\in\mathbb{R}^{N\times N},B\in\mathbb{R}^{N\times 1},C\in\mathbb{R}^{1\times N}$, $h_t$ is $N$-dimension vector, $x_t,y_t,\Delta$ are numbers. Let $A$ be diagonal, $A$ can be also be storaged in $N$-dimension vector. Thus, considering all dimensions, the data structure is $A\in\mathbb{R}^{D\times N},B\in\mathbb{R}^{D\times N},C\in\mathbb{R}^{D\times N},\Delta\in\mathbb{R}^{D}$.

\subsection{Mamba}

Recently, Mamba, a structured state space sequence model with a selection mechanism and scan module (S6), has emerged as a powerful tool in sequence modeling tasks. In S4, $A,B,C,\Delta$ is time-invariant at any time $t=1,2,...,L$. However, Mamba introduces a novel selection mechanism that allows the model to dynamically choose which parts of the input sequence are relevant for making predictions. This mechanism helps the model focus on important information while ignoring irrelevant or noisy data, leading to improved generalization and performance. In selection mechanism, $B\in\mathbb{R}^{B\times L\times N},C\in\mathbb{R}^{B\times L\times N},\Delta\in\mathbb{R}^{B\times L\times D}$ can be learned from $X\in\mathbb{R}^{B\times L\times D}$ using fully-connected layers. By discretization, for each batch and each dimension, let $x_t,h_t,y_t$ denote the input, hidden state and output at time $t=1,2,...,L$, respectively, S4 model can be written as
\begin{equation}
    h_t=\bar A_t h_{t-1}+\bar B_t x_t
\end{equation}
\begin{equation}
    y_t=C_t h_t
\end{equation}

The scan module operates by applying a set of learnable parameters or operations to each window of the input sequence. These parameters are typically learned during training and can include convolutions, recurrent connections, or other types of transformations. By sliding this window over the entire sequence, the scan module is able to capture patterns and dependencies that span multiple time steps.

\subsection{MambaLithium}

\begin{figure}
    \centering
    \includegraphics[width=0.46\textwidth]{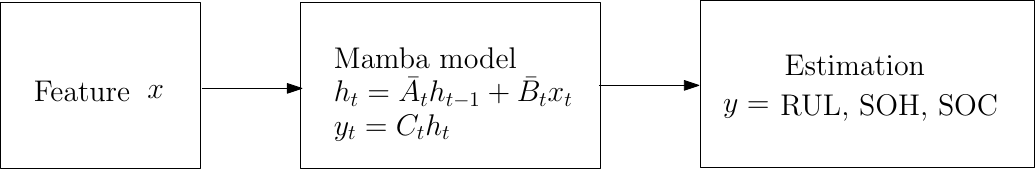}
    \caption{MambaLithium framework}\label{fig:structure}
\end{figure}

The MambaLithium model begins by processing the features mentioned in Section \ref{sec:lithium}. The features are normalized using z-score normalization. Then, the features are fed into the Mamba model with $N=16$, to capture temporal dependencies and extract relevant information. The Mamba model, through its internal mechanisms, is able to effectively mine the patterns and relationships within the input data. The output of the Mamba model is then reduced to a one-dimensional representation for regression prediction. To train the model, the L1-loss is chosen as the loss function, as it measures the average difference between the predicted and actual values. By minimizing this loss, the model aims to improve its accuracy in predicting RUL, SOH and SOC. The framework of MambaLithium is shown on Fig. \ref{fig:structure}. 

The experiments are on an NVIDIA GTX3060 GPU with 12GB memory. The model is trained through Adam optimizer \cite{2014Adam}. The epoch number is 100 and learning rate is 0.01.

\section{Experiments}

\subsection{RUL and SOH estimation}

This paper conduct empirical studies of RUL and SOH estimation on the data following \cite{kong2021voltage,wen2024physics}. The datasets are defined as CaseA and CaseB. Root of mean square error (RMSE) is adopted as evaluation metric.
\begin{equation}
    RMSE = \sqrt {\frac{1}{n}\sum\limits_{t = 1}^n {{{({{\hat X}_t} - {X_t})}^2}} }
\end{equation}
Here $\bar X_t$ denotes the mean value of $X_t$. Lower RMSE denotes better performance. 

\begin{figure}
    \centering
    \includegraphics[width=0.46\textwidth]{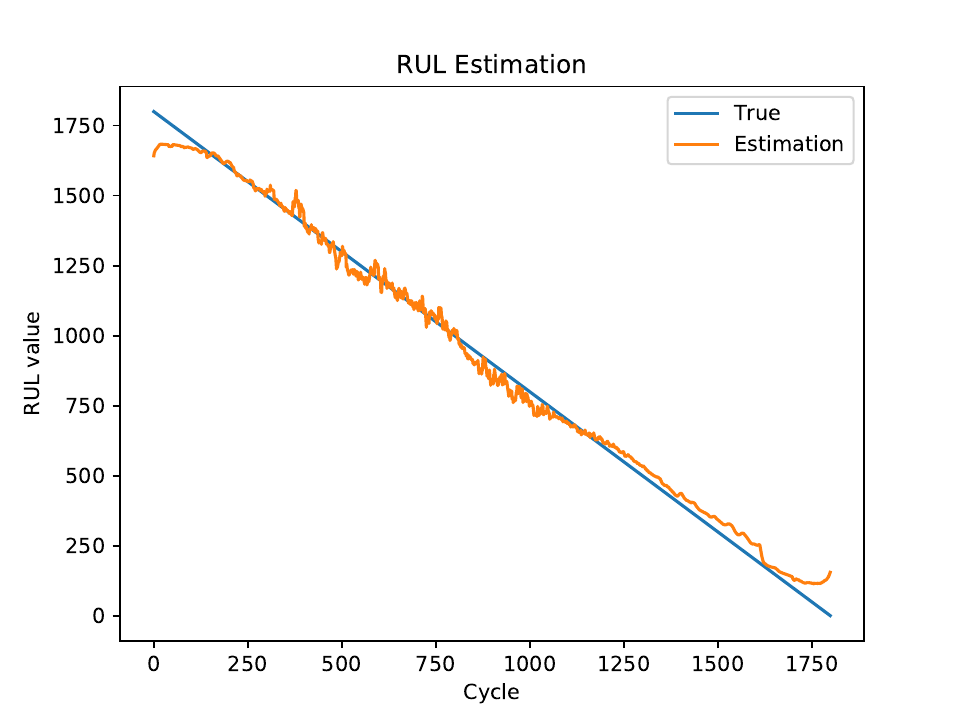}
    \caption{Curve on CaseA of RUL estimation}\label{fig:stock1}
\end{figure}

\begin{figure}
    \centering
    \includegraphics[width=0.46\textwidth]{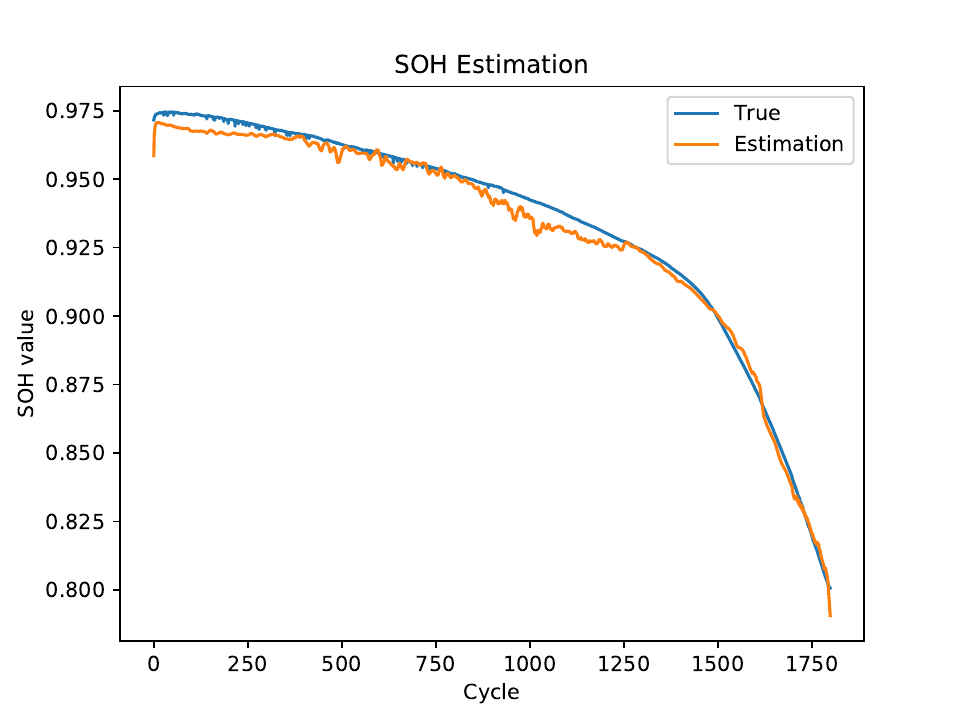}
    \caption{Curve on CaseA of SOH estimation}\label{fig:stock2}
\end{figure}

The recent state-of-the-art (SOTA) method \cite{wen2024physics} is adopted as the baseline. The RUL estimation results are shown in Table \ref{tab:stock1}. The estimated RUL curve of CaseA is shown in Fig. \ref{fig:stock1}. The SOH estimation results are shown in Table \ref{tab:stock2}. The estimated SOH curve of CaseA is shown in Fig. \ref{fig:stock2}. MambaLithium is superior to baseline model. The results of MambaLithium demonstrate its ability to accurately estimate RUL and SOH. The utilization of the Mamba framework allowed the model to capture temporal dependencies and extract relevant information effectively.

\begin{table}[]
    \centering
    \caption{Results on RUL estimation (Unit: cycle)}
    \label{tab:stock1}
    \begin{tabular}{@{}ccc@{}}
    \toprule
    Model       & CaseA    & CaseB\\ \midrule
    Baseline & 45.86 & 56.29 \\
    MambaLithium & 39.71 & 47.62 \\\bottomrule
    \end{tabular}
\end{table}

\begin{table}[]
    \centering
    \caption{Results on SOH estimation (Unit: \%)}
    \label{tab:stock2}
    \begin{tabular}{@{}ccc@{}}
    \toprule
    Model       & CaseA    & CaseB\\ \midrule
    Baseline & 0.47 & 0.42 \\
    MambaLithium & 0.40 & 0.38 \\\bottomrule
    \end{tabular}
\end{table}

\subsection{SOC estimation}

This paper conduct empirical studies of SOC estimation on the data following \cite{yang2019soc,chen2024lstm}. There are three datasets for SOC prediction: DST, FUDS, US06. Either of these datasets can be testing set, and the corresponding other two datasets are used for training. The SOC are measured under cold temperature (average about 0${}^\circ$C), room temperature (average about 25${}^\circ$C) and hot temperature (average about 50${}^\circ$C). Root of mean square error (RMSE) is adopted as evaluation metric.

\begin{figure}
    \centering
    \includegraphics[width=0.46\textwidth]{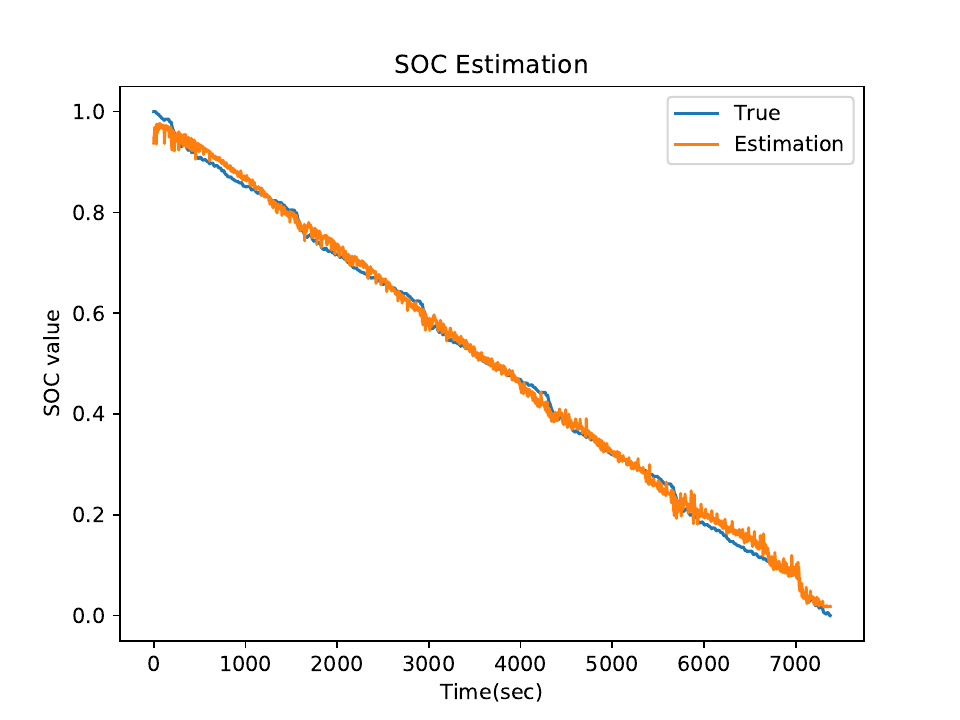}
    \caption{Curve on FUDS dataset of SOC estimation under room temperature (average about 25${}^\circ$C)}\label{fig:stock3}
\end{figure}

The recent state-of-the-art (SOTA) method \cite{wen2024physics} is adopted as the baseline. The SOC estimation results are shown in Table \ref{tab:stock3}. The estimated SOC curve of FUDS dataset under room temperature (average about 25${}^\circ$C) is shown in Fig. \ref{fig:stock3}. MambaLithium surpasses the baseline model, exhibiting a remarkable capacity to precisely estimate the SOC. Leveraging the powerful Mamba framework, the model adeptly captures temporal dependencies and efficiently extracts pertinent information, thereby enhancing its overall performance.

\begin{table}[]
    \centering
    \caption{Results on FUDS dataset of SOC estimation (Unit: \%)}
    \label{tab:stock3}
    \begin{tabular}{@{}cccc@{}}
    \toprule
    Model       & 0${}^\circ$C    & 25${}^\circ$C & 50${}^\circ$C\\ \midrule
    Baseline & 2.98 & 1.78 & 2.04 \\
    MambaLithium & 2.50 & 1.45 & 1.95 \\\bottomrule
    \end{tabular}
\end{table}

MambaLithium is also evaluated when adopting DST or US06 under room temperature (average about 25${}^\circ$C) as testing set, respectively. Table \ref{tab:stock4} shows the results. MambaStock has demonstrated high accuracy on different datasets.

\begin{table}[]
    \centering
    \caption{Results on DST and US06 dataset of SOC estimation under room temperature (average about 25${}^\circ$C) (Unit: \%)}
    \label{tab:stock4}
    \begin{tabular}{@{}ccc@{}}
    \toprule
    Model       & DST    & US06\\ \midrule
    MambaLithium & 1.62 & 0.95 \\\bottomrule
    \end{tabular}
\end{table}

\section{Conclusions}

Lithium-ion batteries have recently emerged as a crucial component in the realm of electric vehicles and new energy industry. Their operational excellence hinges significantly on three fundamental states: remaining-useful-life (RUL), state-of-health (SOH), and state-of-charge (SOC). Given the remarkable prowess of Mamba (Structured state space sequence models with selection mechanism and scan module, S6) in sequence modeling tasks, this paper introduces MambaLithium, a state-of-the-art selective state space model tailored specifically for the precise estimation of these pivotal battery states. By harnessing the sophisticated algorithms of Mamba, MambaLithium adeptly captures the intricate aging and charging behaviors of lithium-ion batteries. By focusing on critical states within the operational range of the battery, MambaLithium not only enhances estimation precision but also maintains computational stability. Rigorous experiments conducted using real-world battery data have confirmed the model's preeminence in predicting battery health and performance indicators, surpassing current techniques. The proposed MambaLithium framework holds immense promise for enhancing battery management systems and fostering sustainable energy storage solutions.



\bibliographystyle{IEEETrans}
\bibliography{reference}


\vfill

\end{document}